# Historical Sunspot Observations: A Review

J. M. Vaquero
Departamento de Física Aplicada, Universidad de Extremadura, Cáceres, Spain [E-mail: jvaquero@unex.es]

**Abstract**

Early observations of sunspot were realised by the naked eye. Possible utilization of these records for studying the long-term change in the Sun is discussed here. Other historical sunspot observations with camera obscuras are also discussed. Moreover, the best record of the behaviour of the Sun exists for the last four centuries thanks to the observations of sunspots with telescope. These observations should allow us to know the number, position, and area of sunspots as well as some relevant episodes (Maunder Minimum, optical flares, etc.). Rudolf Wolf developed the first reconstruction of solar activity in the 19th century. The next reconstruction was made by Hoyt and Schatten in 1998 by improving the database and using a new methodological approach. Here some mistakes, pending tasks and minor improvements are discussed.

## 1. Introduction

Knowledge of solar activity during past centuries is of great interest for many purposes. Long series of solar variability would be interesting for geophysicists and climatologists because of the influence the Sun has on our planet and for solar physicists because they would provide a better understanding of the Sun itself.

Historical documents can help us knowing about the behaviour of the Sun during the last centuries. In these documents, we can find direct observations of the Sun such as measurements by early astronomers, observations of naked-eye sunspots, and eclipses. Also we can find documents providing historical information about the Earth's magnetic field (geomagnetic declination) or auroral observations. On longer time scales, other procedures can also be used for the reconstruction of solar activity such as cosmogenic radionuclides (Usoskin and Kovaltsov, 2004).

The aim of this paper is to present our knowledge of the Sun's behaviour during the last centuries from the observations of sunspots, and to show that this is a field that can be fruitfully developed because there is information of interest is still buried in archives and libraries.

## 2. Naked-eye sunspots

It is occasionally possible to observe a sunspot by eye (Schaefer, 1993) when the spot is sufficiently large and there are certain atmospheric conditions (mist, dust, smoke, ...) which reduce the intensity of the sun's light. A great number of historical naked eye sunspots are reported in Oriental historical sources (China, Korea, and Japan). Others are reported in European, Arabic (Vaquero and Gallego, 2002), Indian (Malville and Singh, 1995), and Maya (Burland, 1958) sources. There are some previous papers about the early observation of sunspots. For example, Goldstein (1969) showed that some medieval reports of Venus and Mercury transits could be interpreted as sunspots, and Sarton (1947) compiled some Occidental observations.

There are various catalogues of naked-eye sunspot observations available in the literature (Wittmann and Xu, 1987; Yau and Stephenson, 1988). Wittmann (2003) compiled a record of naked-eye and large sunspots until the present. In the last twenty years, there have also appeared some new records (Křivský, 1985; Stephenson and Willis, 1999; Vaquero and Gallego, 2002; Moore, 2003; Lee et al., 2004). Vaquero (2004b) reported a naked-eye sunspot observed by Galileo in 1612 (Galileo, 1613). It is intriguing that it has been forgotten until now by the compilers of the naked-eye sunspots catalogues in spite of Galileo's fame. It would seem likely therefore that a systematic search in Occidental sources could well increase the list of these records.

Some visibility criteria have been presented by Keller and Friedli (1992) and Schaefer (1993), and modern systematic observations have been reported by Mossman (1989), Heath (1994), and Wade (1994). The reliability of the naked-eye observations of sunspots has been assessed by comparing carefully the Oriental sunspot sightings from 1862 onwards with contemporary Occidental white-light images of the Sun (Willis et al., 1996). These observations might be related to variations in some meteorological parameters (Willis et al., 1980; Eddy et al., 1989; Scuderi, 1990; Hameed and Gong, 1991), such as the number of days of mist or of dust storms. Nevertheless, some authors have reconstructed time series of annual number of naked-eye sunspot in spite of the evident climatic contamination of the series.

Nagovitsyn (2001) and Vaquero et al. (2002a) used catalogues of naked-eye observations of sunspots to construct time series showing the behaviour of the solar activity during the last two millennia. Figure 1a was constructed by calculating the 50-year moving average of the series. The strong 250-year period reported in Vaquero et al. (2002a) now stands out far more sharply, as do the known minima of Oort, Wolf, Spörer, Maunder, and Dalton. The Maunder and Dalton Minima are somewhat displaced from their original positions due to the moving average procedure. One also observes the Medieval Maximum centred in the first half of the 12th century. Thus, the series of the annual number of naked-eye sunspots agree qualitatively with other proxies of solar activity (Vaquero et al., 2002a; Ogurtsov et al, 2002). Figure 1b shows a comparison between the 50-year moving average of annual naked-eye sunspots and Group Sunspot Numbers. It is evident that, from a quantitative point of view, the historical naked-eye observations must be interpreted with caution (Willis et al., 1996; Usoskin and Kovaltsov, 2004). Thus, figure 1a does not pretend to represent a quantitative variation of solar activity. Naked-eye observation cannot be interpreted quantitatively because only observed "spots" are reported and thus we can not distinguish between "no-spot" and no-observation cases. Moreover, only a 67 per cent of Oriental naked-eye observations have been confirmed by telescopic observations during the period 1874-1918 (Willis et al., 1996).

Are the historical naked-eye observations useful for space climate studies then? The high resolution of these observations, compared with other proxies, can be used in a coherent form by space climate researchers. Willis and Stephenson (2001) show evidence for an intense recurrent geomagnetic storm during December in 1128 AD using historical aurorae and naked-eye sunspot observations. Moreover, Willis et al. (2005) recently used a comprehensive collection of catalogues of ancient sunspot and auroral observations from East Asia to identify possible intense historical geomagnetic storms in the interval 210 BC–1918 AD.

## 3. Sunspots observations with Camera Obscuras

Spread of the telescopic instrumentation during the early 17th century was a revolution in astronomy, and led to the discovery of sunspots (Bray and Loughhead, 1964; Brody, 2002). However, another important scientific instrument was very useful for astronomers already in the 16th and early 17th century: camera obscura. Small camera obscuras were used from very early times, but during the 16th century until the 1750s meridian lines in major cathedrals were used to determine the Sun's position and the Earth's orbit (Heilbron, 1999). As an example of its astronomical use, Sigismondi and Fraschetti (2001) presented five measurements of the solar diameter made by Tycho and Kepler with pinhole instruments, reduced and published by Kepler (1604). Obviously, sunspots were also observed with the camera obscura. There is a famous text of Galileo where he invites persons who do not want to use the telescope to use a broken window of a temple as a camera obscura to observe sunspots.

The first observation known till now of a sunspot through a camera obscura was done by Kepler. During the months of April and May of 1607, Kepler had been engaged in observing the planet Mercury in the evening sky. Contemporary planetary ephemerides predicted that Mercury would be in conjunction with the Sun. On 28 May, Kepler observed the Sun with a camera obscura. He detected a small spot on the solar disk and was convinced that he was observing a transit of Mercury. He published his discovery in a short treatise on the comet of 1607 and in more detail in his Phænomenon singulare (Kepler, 1609). A few years later, Kepler realized that his 1607 observation could not have been Mercury but had been a large sunspot (Figure 2).

The earliest drawing, to the best of our knowledge, of a solar observation using camera obscura is due to Frisius (1545). He observed the 1544 solar eclipse from Louvain. No sunspots were recorded. The observation was made during the Spörer Minimum and the probability of seeing a sunspot would have been very low. It is the only available direct observation of the Sun during the late Spörer Minimum (see Figure 2a,b).

The use of small camera obscuras in astronomical observation ceased with the generalization of the use of telescopes. However, during the 17th and 18th centuries, meridian lines in cathedrals in Bologna, Rome, Florence, and Paris served as solar observatories making careful and continuous study of the Sun possible. The earliest of these cathedral observatories was a big church of San Petronio in Bologna. The heliometer of San Petronio consists of two separate pieces. One piece lies on the floor; it is a perfectly horizontal rod running due north for about 67 meters from a spot under one of the side chapels to the front door of the church. The other part is a small hole 2.5 cm in diameter set in a horizontal metal plate fixed in the roof of a chapel. The hole is permanently open so as to give free access to the sun's rays around noon throughout the year. For eighty years (1655-1736), some scientists and clerics observed the Sun, filling about 300 pages in the register published by Eustachio Manfredi (1736). Each entry includes a description of the weather, the distances of the sun's limbs from the vertex corrected for the penumbra, and the apparent diameter of the sun, all given to seconds of arc.

Though one must bear these camera obscura observations in mind, the great majority of sunspot observations were made with telescopes.

## 4. The sunspot numbers

Two major reconstructions of solar activity have been made on the basis of direct observations of the Sun. The International Sunspot Numbers (Wolf or Zürich sunspot numbers) have long served as the primary time series defining solar activity since the year 1700. This time series was derived by Rudolf Wolf in the 19th century and has been maintained by his successors (Waldmeier, 1961; McKinnon, 1986, Vanlommel et al., 2004). Wolf defined the sunspot number, $R_Z$, as

$$R_Z = k(10g + f), \qquad (1)$$

where g is the number of sunspot groups, f is the number of individual sunspots, and k is a correction factor for each observer. The daily $R_Z$ value is calculated by using only the input from one observer selected as the primary observer for the given period. If the primary observer could not make an observation, then secondary, tertiary, etc., observers are used to complete the maximum possible number of days. For years up to 1817, the number of missing days was so great that Wolf only tabulated monthly means. There are no observations for many months from 1749 to 1818, and for a few months after 1818. Wolf filled in these months by interpolation and using magnetic needle observations. Thus, the numbers $R_Z$ are a mixture of direct sunspot observations and calculated values.

Hoyt and Schatten (1998 - HS98 hereafter) made a new reconstruction of solar activity from sunspot observations. This time series is known as the Group Sunspot Number, $R_G$, because it uses the observed number of sunspot groups. Hoyt and Schatten defined the Group Sunspot Number as

$$R_G = \frac{12.08}{N}\sum k_i' G_i \qquad (2)$$

where $G_i$ is the number of sunspot groups recorded by the i-th observer, $k_i$ is the i-th observer's correction factor, N the number of observers used to form the daily value, and 12.08 is a normalization factor chosen to make the mean of the $R_G$'s identical with the mean of the $R_Z$'s for 1874 through 1976. Daily, monthly, and yearly means were derived from 1610 to the present. They calculated daily values of solar activity on 111 358 days for 1610–1995, compared to 66 168 days for the International Sunspot Numbers, tabulating estimates of their random and systematic errors. This series has complete or nearly complete coverage from about 1800 to 1995 and from 1645 to 1727 (Figure 3). From 1610 to 1644 (Figure 3b) and from 1728 to 1799 (Figure 3c), there are many years with only sparse observations. For six years (1636–1637, 1641, 1744 –1745, and 1747) there exist no reports of sunspot observations.

The Group Sunspot Numbers are strongly recommended for analysis of sunspot activity before 1880. The Wolf and Group numbers are not different alternative proxies of solar activity because the Group numbers are an upgrade of the Wolf numbers. However, the old Wolf numbers are still used in several studies especially in the climate community. Examples are the works of Gimeno et al. (2003), Schröder et al. (2004), and Benestad (2005).

Hathaway et al. (2002) examined the Group Sunspot Numbers to determine their utility in characterizing the solar activity cycle. They find that the $R_Z$ numbers follow the 10.7-cm radio flux and total sunspot area measurements only slightly better than the $R_G$ numbers. The "Waldmeier Effect" (the anti-correlation between the cycle's amplitude and the length of the ascending phase of the cycle) and the "Amplitude-Period Effect" (the anti-correlation between the cycle amplitude and the length of the previous cycle from minimum to minimum) are much more apparent in the $R_Z$ numbers. The "Amplitude-Minimum Effect" (the correlation between cycle amplitude and the activity level at the previous minimum) is equally apparent in both the $R_Z$ numbers and the $R_G$ numbers. The "Even-Odd Effect" (in which odd-numbered cycles are larger than their even-numbered precursors) is somewhat stronger in the $R_G$ numbers but with a tighter relationship in the $R_Z$ numbers. And finally, the secular trend of the sunspot numbers is much stronger in Group numbers. Hathaway et al. (2002) concluded that the $R_G$ numbers are most useful for extending the sunspot cycle data further back in time and thereby adding more cycles and improving the statistics. However, the $R_Z$ numbers are slightly more useful for characterizing the on-going levels of solar activity. The spectral characteristics of the $R_Z$ and $R_G$ series are very similar for the 19th and 20th centuries. Some minor differences are noticed for the 18th century (Kane, 2002). Moreover, Vaquero et al. (2006) showed, in studying estimates of the solar cycle length using $R_G$ and $R_Z$, that the use of $R_G$ resolves some doubtful solar cycle length estimates obtained around 1800 using $R_Z$. Figure 5a shows both sunspot numbers. A 25-year moving average of the two indices is presented in Figure 4b, showing the long-term differences between them.

## 4.1. Some mistakes

HS98 performed a great work on solar activity reconstruction from documentary sources. However, it still contains some minor errors and inexactitudes. In the next sub-sections, some calendar problems or the use of only partial information from eclipse observations are reported.

1) Calendar problems

All the observations in the database of reconstructed solar activity must be referred to the same calendar (the Gregorian calendar in this case). Some countries still used the Julian (or Old Style) calendar well into the telescope era. Examples are Great Britain (including the British colonies) up to 1752 and Russia up to 1918. It is easy to confuse the Julian and Gregorian dates. For example, the dates of sunspot observations by Jeremiah Horrox or Horrocks (1619-1641), an English astronomer who made the first observation of a transit of Venus, are erroneously interpreted in HS98. HS98 found 6 observations by Horrox during the year 1638 in Liverpool (England), using the Opera posthuma of Horroccii (1673). The observation dates are 22-24 May (2 sunspot groups) and 20-22 Oct (2 sunspot groups). However, these dates are referred to the Julian calendar. E.g., the Venus transit of 4 December 1639 was observed by Horrox who, however, wrote in his book (Horroccii, 1673, p. 393) the date of 24 November. He also observed the solar eclipse on 1 June 1639 but wrote in his book (Horroccii, 1673, pp. 387-389) the date 22 May 1639. It is evident that he used the Julian calendar (ten days of difference with respect to the Gregorian calendar in that epoch).

2) One methodological inconsistency

HS98 deviated from their reconstruction methodology on just one occasion. They found in the Cambridge University Library (Flamsteed paper collection) a letter by William Crabtree (1610-1644) with information on sunspots. They wrote in their Bibliography: "According to a letter by Crabtree the average number of spot groups seen in 1638 and 1639 were 4 to 5 per day. The database has Greenwich fill values to give 4 to 5 groups per day. This substitution technique was used to simplify the analysis. This is the only place in the entire database where we do this type of substitution". A comparison between the observed (by Horrox and Gassendi) and the estimated (from Crabtree's comment) values from HS98 is shown in Figure 5. The number of observed groups is smaller than the number of estimated groups.

3) Groups or sunspots?

HS98 corrected some of Wolf's estimates (e.g., Horrebow) since Wolf had not differentiated between groups and spots. Equally, some records of HS98 may suffer from this problem.

Here we propose an example of the observation of the solar eclipse on 15 July 1730. According to HS98, the observer is Hallerstein from Pekin. However, we know that although Hallerstein (1768) collected observations in his book, he was not an observer. The observers were the Jesuits I. Kegler and A. Pereira (1733, 1799). In the description of the contacts of the Moon with the sunspots, the observers describe seven spots on the solar disc. HS98 took this to be 7 groups of sunspots. Nevertheless, a close reading of the observation shows that these seven spots can easily be grouped into three groups. The first group consists of just the largest sunspot observed near the solar limb. The second group consists of 4 spots in the solar northwest, and the third consists of two spots in the southwest. Thus, the number of groups should be reduced from 7 (HS98) down to 3. This might partly explain the anomalously high number of groups during 1730, because in that year there were very few observations.

4) Change of date

Another problem is the change of dates (year or month). As an example, HS98 picked up an observation of J. Huxham published in the Philosophical Transactions describing the Venus transit in 1736 (Huxham, 1744). However, the year of observation given in HS98 is 1739, creating a spurious observation.

5) Partial information from eclipse observations

A considerable part of the information in the work of HS98 about sunspots during the 18th century comes from observations of solar eclipses. An usual task undertaken by astronomers of the 18th century was to time the moments of apparent contact between the Moon and the sunspots during the progress of the eclipse. The Moon covers all the spots on some occasions, and on others not. Therefore, to use the sunspots that appear in a list of times of contact during eclipses can be dangerous because it can underestimate the number of sunspots and groups.

An example of this situation is the observation of a solar eclipse in Lyon by French Jesuits on 25 September 1726. HS98 list this observation among the observations of Souciet (1729), though that author was only a compiler of the work. In this case, we can read 4 occulted sunspots in the table of contacts of the eclipse, and, indeed, HS98 took 4 groups of sunspots, one group for every occulted sunspot. Nevertheless, one can see from the drawing that the observers provided in the report (Figure 6) that 6 spots were observed on the solar disc. Spots C and D (occulted by the Moon) are in the penumbra itself and have to be considered as one group of sunspots. Therefore, one must consider 5 groups of sunspots from this observation of the eclipse of the Sun of 25 September 1726.

## 4.2. Tasks remaining after Hoyt and Schatten's work

There are some pending tasks in the comments on the observers in the Bibliography of HS98. One could classify these tasks into six categories (see Table I):

1) Lost original observations. Many manuscripts are lost. In some cases, our knowledge of the observations is because they were copied by other persons. Discovery of the original observations would allow one to verify the information. In other cases, the observations have been lost, and locating them would help to improve notably our knowledge of past solar activity.

2) Lost daily data. In some cases, we only know monthly or annual averages from some observers. Locating the daily information would allow us to improve the database of observations.

3) Observations not included in the database. For some reason, some published information was not incorporated into the database compiled by HS98.

4) Observations too vague, ambiguous or narrative style. On occasions, observational reports are written in a narrative form that makes it very difficult to construct trustworthy tables with that information.

5) Incomplete observations. On other occasions, HS98 obtained information of an observer knowing that there existed other volumes of observations that were not accessible to them. Retrieval of the remaining observations would be a simple task for those persons who do have access to the complete collection.

6) References need to be re-checked. On some occasions, HS98 themselves note a failure to annotate the correct reference of books or manuscripts.

### 4.3. Small improvements

Undoubtedly, a great problem in reconstructions from historical solar observations is the influence that the days with no available information have on the monthly and annual averages. Some authors have treated this problem by trying to fill the gaps in the series (Letfus, 1999) or by developing statistical methods that estimate the margin of error of the average even if the number of observations per year is very small (Usoskin et al., 2003a).

Another possibility is to find new observations to fill up the gaps in the series. For example, Vaquero (2003) describes observations of an Italian astronomer Andrea Argoli (1570-1657). Argoli recorded in his book Pandosion sphaericum (Argoli, 1653) that he had not seen any sunspot in 1634 from 19 July until mid September. During this period the French astronomer Gassendi also observed the Sun with the same result. Therefore, Argoli's observations do not fill any gap in the series, but only confirm the information that we already have.

Another example of solar observations, performed by a Mexican astronomer J. A. Alzate during the year 1784, was analysed by Vaquero (2004a). These observations are very valuable for the reconstruction of solar activity because HS98 only found five observations during that year – all performed by J. C. Staudacher. Using conjointly the data provided by Alzate and Staudacher for 1784, one can determine a value of $R_G$ equal to 0.3±0.1 with eighty records for that year (HS98 value: 4.8). Vaquero et al. (2005) presented a “lost” sunspot observation made by a Portuguese scientist Sanches Dorta during his observation of the solar eclipse of 9 February 1785 from Rio de Janeiro (Brazil). This record was not included in the database compiled by HS98. A value of $R_G$ equal to 16.4±3.4 for the year 1785 is obtained (HS98 value: 16.0).

## 5. Position and area of sunspots

The position and area of sunspots are important parameters for the long term study of the Sun together with the number of sunspots. We can then study the long term evolution of the solar rotation, the asymmetry between the two hemispheres, the Sun’s active longitudes, and a long list of other questions. There are some sporadic measurements of positions and areas of spots in the 18th century. Nevertheless, the information most widely used in the current studies comes from the Royal Greenwich Observatory (RGO). The RGO compiled sunspot observations from a small network of observatories to produce a dataset of daily observations starting in May 1874. The observatory concluded this dataset in 1976 after the US Air Force (USAF) started compiling data from its own Solar Optical Observing Network (SOON). This work was continued with the help of the US National Oceanic and Atmospheric Administration (NOAA) with much of the same information being compiled through to the present (http://solarscience.msfc.nasa.gov/greenwch.shtml).

However, several solar physicists had observed the sunspot positions or areas prior to 1874. Dr Soemmering made this kind of observation during the years 1826-1829, but his work is now lost. Some information about the Soemmering observations can be obtained from Carrington (1860b). Carrington (1863) obtained sunspot positions from November 1853 to March 1861, and Peters (1907) at Hamilton College from 1860 to 1870. Other data prior to the RGO observations come from Spörer (Wöhl and

Balthasar, 1989).

Also before the measurement program of the RGO, De la Rue, Stewart and Loewy (1869, 1870) made a great effort in observing positions and areas with the Kew Photoheliograph during the years 1862-1866. They also compiled drawings and photographs of the sunspots in the period 1832–1868. From these drawings and photographs, they determined fortnightly values of the sunspot areas. The measurement method and the reliability of this series have been studied by Vaquero et al. (2002b). Some reconstructions of sunspot areas are available (Vaquero et al., 2004; Nagovitsyn et al., 2004; Nagovitsyn, 2005).

Balthasar et al. (1986) showed that the RGO data are of great interest for determining the differential rotation of the Sun on the basis of the complete sample (1874-1976). Also, Wöhl and Balthasar (1989) compared the solar rotation velocity inferred from the RGO data and from Spörer's data for the years 1883-1893, finding a difference of the order of 0.1 deg/day. However, an analysis of the rotation velocities of only the stable sunspots, which are covered sufficiently by observations in both sets of data, yields no significant differences. The explanation of the 0.1 deg/day difference is the presence of more short-lived sunspots in the RGO, which show a more rapid rotation velocity.

## 6. Episodes

During the 400 years of systematic sunspot observations with telescopes, several episodes of special interest have been recorded. Probably, the most interesting was the Maunder Minimum (Eddy, 1976, 1986; Ribes and Nesme-Ribes, 1993; Usoskin et al., 2001). However, other interesting events can be indicated too. Great activity complexes have been recorded during the last centuries. Probably, the best compilation was made by Wittman (2003). The observations of white-light flares may be the most striking of this kind of observation. Also, some other events stand out. For example, the extraordinary extent of cycle #4 has constituted a puzzling issue for the scientific community (Sonett, 1983). Usoskin, Mursula, and Kovaltsov (2001) suggest that solar cycle #4 was in fact a superposition of two cycles: a normal 10-year long cycle between 1784 and 1793 followed by a short and weak cycle in 1793-1800. Arguments for and against the “lost cycle” theory have been proposed in recent work (Usoskin, Mursula and Kovaltsov, 2002, 2003b; Krivova, Solanki and Beer, 2002).

In the last few years, numerous works have appeared on the Maunder Minimum, including major reviews by Soon and Yaskell (2003) and more recently by Miyahara, Sokoloff, and Usoskin (2006). Another recent work has considered the solar rotation during the 17th century (Casas et al., 2006), using observation of a sunspot by Nicholas Bion. Table II compares the observations of Bion with those of other astronomers during the same period. Surprisingly, during the October 1762 observation period, other astronomers who had observed the Sun, some of them (Hevelius and Picard) probably using adequate apertures, did not note the presence of sunspots. The case of Montanari is special because he was not using a telescope but the camera obscura of San Petronio in Bologna. Indeed, 2218 observations made with the San Petronio camera obscura were used in HS98 to reconstruct the solar activity during the Maunder Minimum (Table III) from the book published by Manfredi in 1736. In their paper published in 1996, Hoyt and Schatten established clearly that the Sun was monitored correctly during the Maunder Minimum and that the absence of sunspots is beyond doubt. Nevertheless, examples like Bion's observation indicate to us that we must continue working at the fine details of sunspot activity that could improve our knowledge of the Sun during the Maunder Minimum.

White-light flares (WLFs) are major flares in which small parts of the Sun become visible in white light. Such flares are usually strong X-ray, radio, and particle emitters. Neidig and Cliver (1983) compiled a catalogue of 57 WLFs reported between 1859 and 1982. The first observation of a white-light flare was found by Hoyt and Schatten (1996). On 27 December 1705, Stephen Gray of Canterbury recorded in a manuscript (in the Flamsteed collection, Cambridge University Library) that he saw a “flash of lightning” near a sunspot. This is one of the indicators of the ability of solar observers during the Maunder Minimum shown by Hoyt and Schatten (1996).

The next white-light flare was observed 154 years later. There are some recent reviews of this Carrington (1860a) and Hodgson (1860) flare observed on 1 September 1859 (Cliver and Svalgaard, 2004; Cliver, 2006; Tsurutani et al., 2003). Here, I show a record recovered thanks to SIGN, a Portuguese project of recovery of meteorological and geophysical information. Table IV shows the monthly values of geomagnetic declination measured at Lisbon during the year 1859 (Observatorio do Infante D. Luiz, 1863). Unfortunately, no daily data are available from Lisbon for that epoch. The greatest variations correspond to the months of August and September. The maximum and minimum values during September were measured on the 2nd day showing clearly the effect of Carrington's flare on the geomagnetic field measured at Lisbon. The extreme values of August were measured on the 28th and 29$^{th}$ days. Also we must emphasize that 19 May was also very active, although to a smaller degree than 2 September.

## 7. Conclusions

We have reviewed historical evidence concerning the number, positions, and areas of sunspots during the last centuries. From this data, various authors have extracted some very important results for astronomy and geophysics in general, and for space climate in particular, about the long-term variation in solar activity. Table V lists the Decalogue of Hathaway and Wilson (2004) on the most important results for space climate from the sunspot record.

As a final comment, I would like to say that we must make an effort to incorporate the historical information on solar activity into studies of the position of sunspots (differential rotation, active longitudes, north – south asymmetries, …) and continue with the effort that has been made during the last years to improve the time series of sunspot numbers from historical observations.

## Acknowledgements

The author would like to thank Drs M. C. Gallego, J. A. García, F. Sánchez-Bajo, R. M. Trigo and M. Vázquez for their comments. This work has made use of the Group Sunspot Number provided by the NOAA/ National Geophysical Data Center and International Sunspot Number provided by Solar Influences Data Analysis Center. It has also made use of the Astrophysics Data System (NASA). All the historical materials used in this work were consulted at the Biblioteca Nacional de Portugal (Lisbon, Portugal) and Biblioteca del Real Observatorio de la Marina (San Fernando, Spain). The data from Lisbon was obtained through the project SIGN (Signatures of environmental change in the observations of the Geophysical Institutes) under POCTI/CTA/47803/2002 project. Support from the Universidad de Extremadura, Gulbenkiam Foundation-Biblioteca Nacional de Portugal, and Junta de Extremadura is gratefully acknowledged.

**Appendix: Historical Sources**

Table I. Annotations expressing caveats concerning the sunspot observations made by various observers from the "Bibliography of Sunspot observers with comments" (Hoyt and Schatten, 1998).

| Annotation | Observers |
|---|---|
| Original observations are lost | Roemer, J. L. Alischer, Williamson of Cheetham Hill, F. Buser, J. G. Fink, H. Flaugergues, M. Fogel, P. Heinrich, J. Plantade, H. B. Rumrill, J. Schaller, C. Scheiner, H. Siverus, Soemmering. |
| Daily data lost | P. Carl, E. Loreta, A. Moisseiv, M. Moye, S. Teslya, V. Tshernov. |
| Observations not included in the database | Anger, Brorsen, Godlee Obs., Sohn, A. Vagetius, Zagreb Observatory. |
| Observations too vague, ambiguous or narrative style | E. F. F. Chladni, Colla, N. von Konkoly, Maclead, Quetelet, Sandt, von Jahn. |
| Incomplete observations | British Astron. Assoc., Mount Holyoke College, S. M. J. Prantner, San Miguel Obs., Taipei Obs. |
| References need to be re-checked | Cassella, Eclipse observers (Roma, 1734), P. B. G. Fontana, P. A. B. Gesu, N. Hartsoeker, J. Picard, G. Trautmann. |

Table II. Observations of sunspots during October-November 1672. White = missing data; Black = one sunspot group; grey = no sunspots.

| **Day** | **Bion** | **Picard** | **Hevelius** | **Montanari** | **Petitus** |
|---|---|---|---|---|---|
| 18 October | Black | White | Grey | White | White |
| 19 October | Black | Grey | White | White | White |
| 20 October | Black | Grey | White | White | White |
| 21 October | Black | White | White | Grey | White |
| 22 October | Black | White | Grey | White | White |
| 23 October | Black | Grey | Grey | White | White |
| 24 October | Black | White | White | White | White |
| 25 October | Black | White | White | White | White |
| 26 October | Black | White | White | White | White |
| 12 November | Black | Black | White | White | Black |
| 13 November | Black | Black | White | White | Black |
| 14 November | Black | Black | White | White | Black |
| 15 November | White | White | White | White | White |
| 16 November | White | White | Grey | White | White |
| 17 November | White | Black | White | White | White |
| 18 November | Black | White | White | White | White |
| 19 November | Black | White | White | White | White |
| 20 November | Black | Black | White | White | Black |
| 21 November | White | White | White | White | White |
| 22 November | Black | Black | White | White | Black |

Table III. Solar observations during the Maunder Minimum from Manfredi (1736) taken as zero spot days by HS98.

| Observer | Period | Number of records |
|---|---|---|
| J. C. Calcina | 1674 | 20 |
| G. D. Cassini | 1656-1670 | 13 |
| A. Fabrius | 1668-1675 | 90 |
| J. F. Gulielmini | 1675-1696 | 406 |
| D. Manzius | 1673 | 1 |
| F. Maraldi | 1673 | 273 |
| P. Mengoli | 1663-1670 | 66 |
| C. Mezzavacca | 1663-1695 | 6 |
| G. Montanari | 1671-1676 | 107 |
| G. Riccioli | 1655-1661 | 92 |
| V. F. Stancarius | 1696-1702 | 1103 |
| I. Uccelli | 1695-1696 | 41 |

Table IV. Mean, minimum, and maximum values of the geomagnetic declination (West) measured in the "Observatorio do Infante D. Luiz" at Lisbon during the year 1859, giving the complete dates of maxima and minima.

| **Month** | **Mean Values** | | **Minimum value** | | **Maximum value** | |
|---|---|---|---|---|---|---|
| | **08:00** | **14:00** | **Day** | **Value** | **Day** | **Value** |
| January | 21° 33.4' | 21° 40.7' | 26 | 21° 28.7' | 5 | 21° 43.7' |
| February | 21° 30.7' | 21° 40.9' | 22 | 21° 26.9' | 26 | 21° 45.6' |
| March | 21° 28.9' | 21° 40.6' | 4 & 20 | 21° 25.6' | 11 | 21° 43.2' |
| April | 21° 26.6' | 21° 42.0' | 13 | 21° 20.7' | 21 | 21° 45.6' |
| May | 21° 29.4' | 21° 40.6' | 19 | 21° 23.5' | 19 | 21° 48.7' |
| June | 21° 27.3' | 21° 38.6' | 28 | 21° 18.7' | 8 | 21° 41.4' |
| July | 21° 28.4' | 21° 39.4' | 3 | 21° 23.7' | 18 | 21° 46.2' |
| August | 21° 26.9' | 21° 38.6' | 28 | 21° 18.1' | 29 | 21° 57.6' |
| September | 21° 27.7' | 21° 39.6' | 2 | 21° 21.5' | 2 | 21° 55.7' |
| October | 21° 27.8' | 21° 38.7' | 21 | 21° 24.7' | 12 | 21° 51.7' |
| November | 21° 27.9' | 21° 36.0' | 7 | 21° 26.0' | 13 | 21° 42.8' |
| December | 21° 29.5' | 21° 35.3' | 30 | 21° 27.7' | 11 | 21° 39.4' |

Table V. Decalogue of Hathaway and Wilson (2004) on important results for space climate from sunspot records.

| **N** | **Result** |
|---|---|
| 1 | Sunspot cycles have periods of 131±14 months with a normal distribution. |
| 2 | Sunspot cycles are asymmetric with a fast rise and slow decline. |
| 3 | The rise time from minimum to maximum decreases with the cycle amplitude. |
| 4 | Large amplitude cycles are preceded by short period cycles. |
| 5 | Large amplitude cycles are preceded by high minima. |
| 6 | Although the two hemispheres remain linked in phase, there are significant asymmetries in the activity in each hemisphere. |
| 7 | The rate at which the active latitudes drift toward the equator is anti-correlated with the cycle period. |
| 8 | The rate at which the active latitudes drift toward the equator is positively correlated with the amplitude of the cycle after the next. |
| 9 | There has been a significant secular increase in the amplitudes of the sunspot cycles since the end of the Maunder Minimum (1715). |
| 10 | There is weak evidence for a quasi-periodic variation in the sunspot cycle amplitudes, with a period of about 90 years. |

(a)

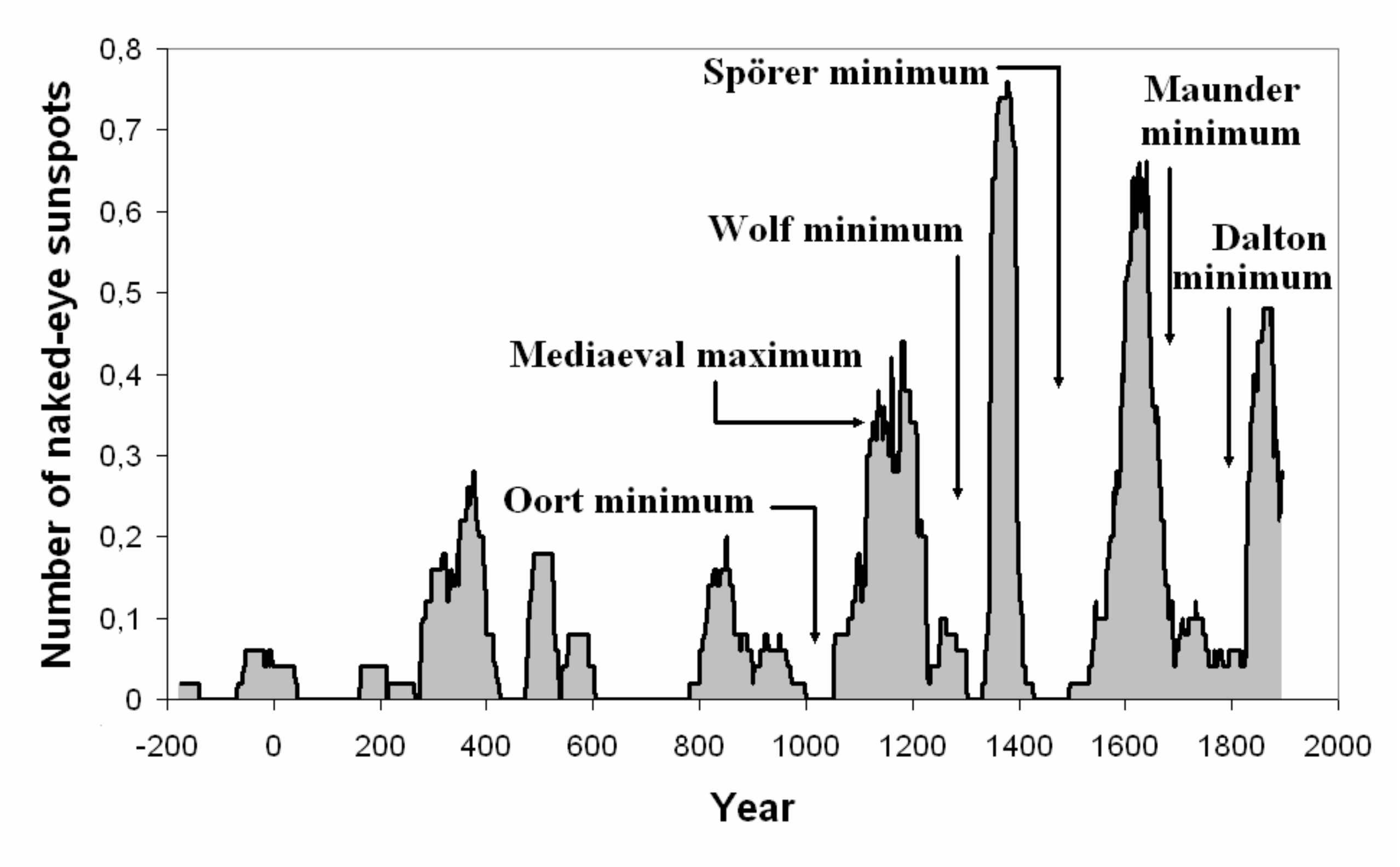

(b)

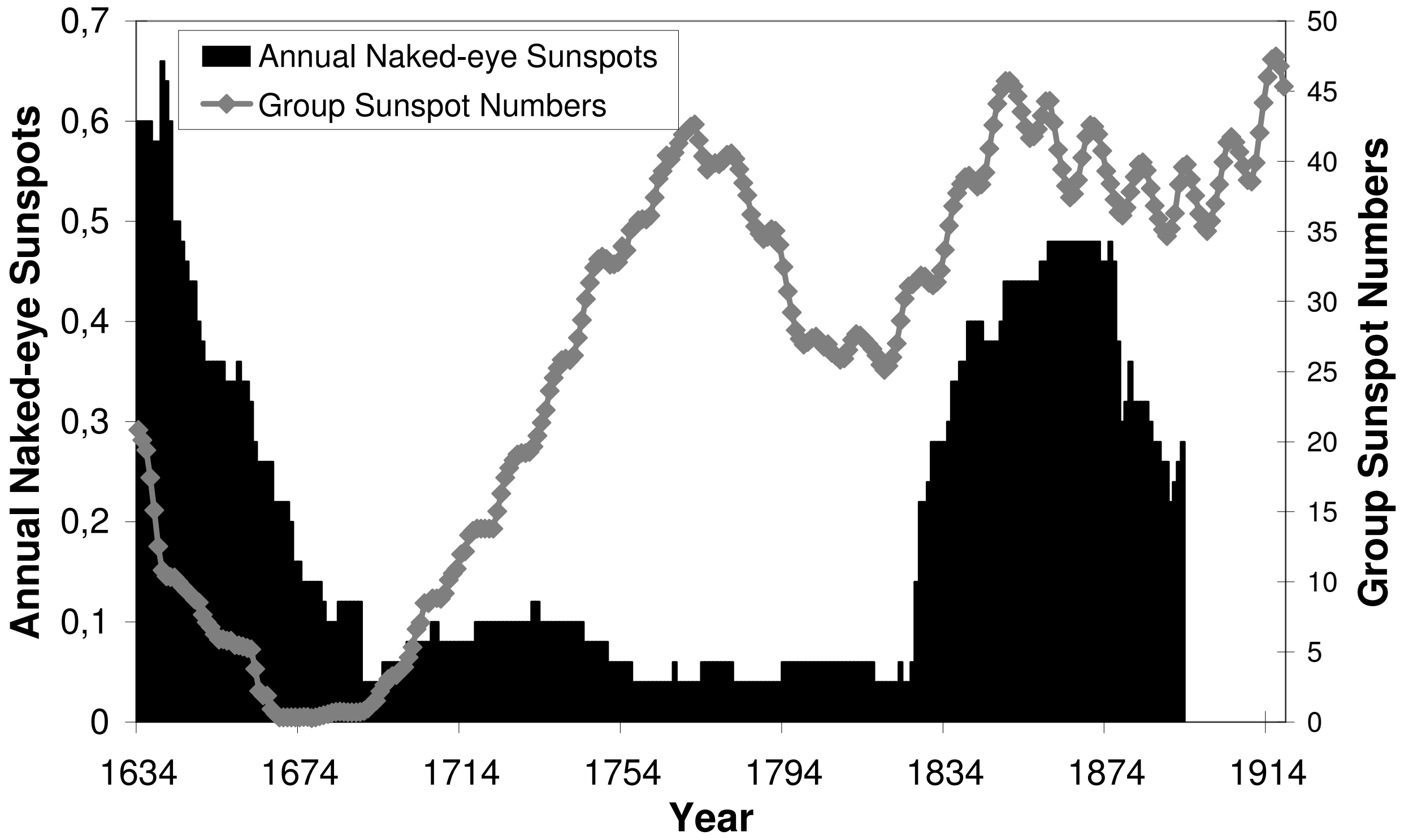

Figure 1. (a) 50-year moving average of the number of annual naked-eye sunspots during the last 22 centuries from Vaquero et al. (2002). (b) A comparison between 50-year moving averages of annual naked-eye sunspots and Group Sunspot Numbers.

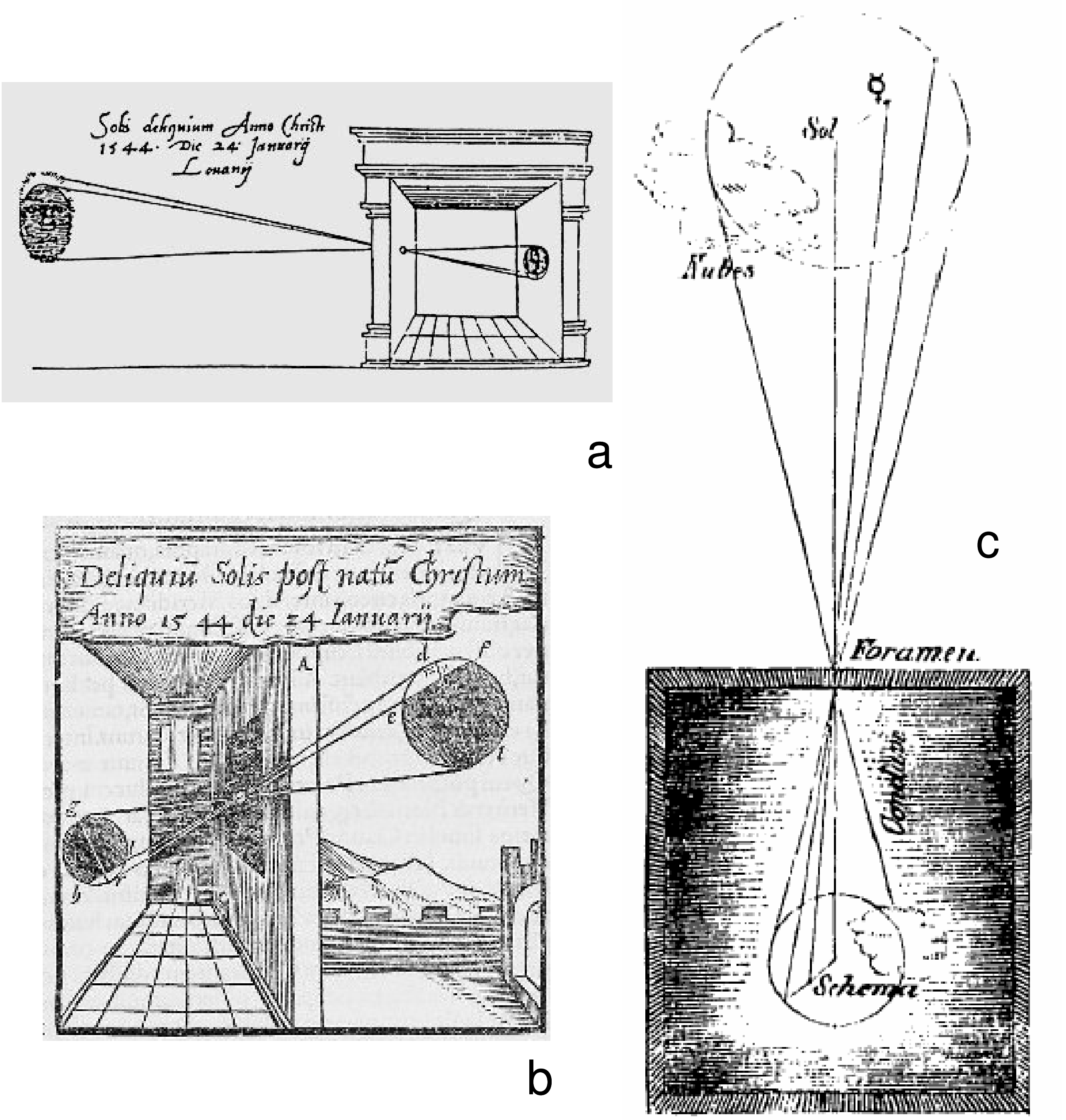

Figure 2. Some early drawings of astronomical use of a camera obscura: (a) 1544 solar eclipse observation by Gemma Frisius from Frisius (1545); (b) the same observation from Regiomontanus (1561); and (c) sunspot observation in 1607 by Kepler (1609).

(a)

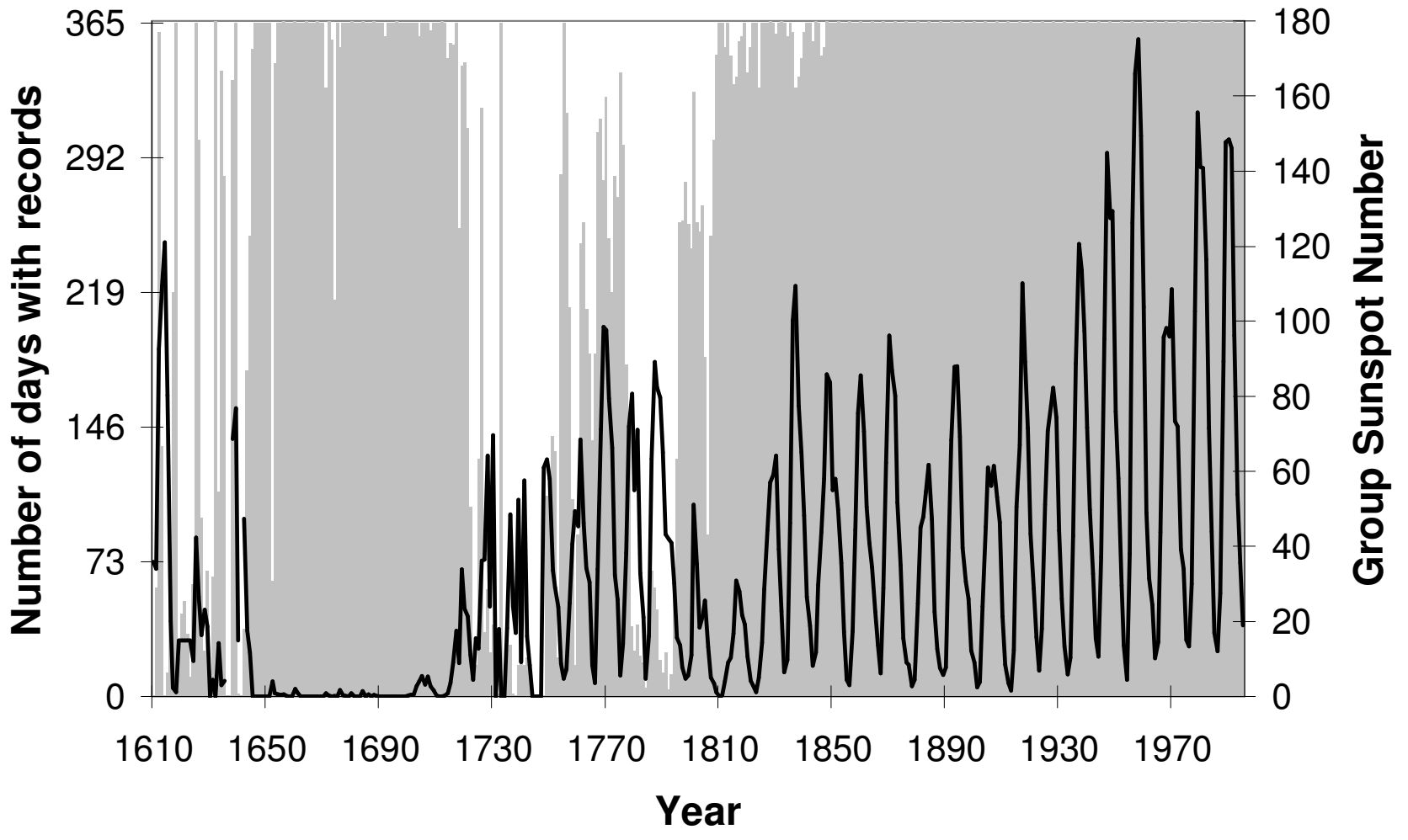

(b)

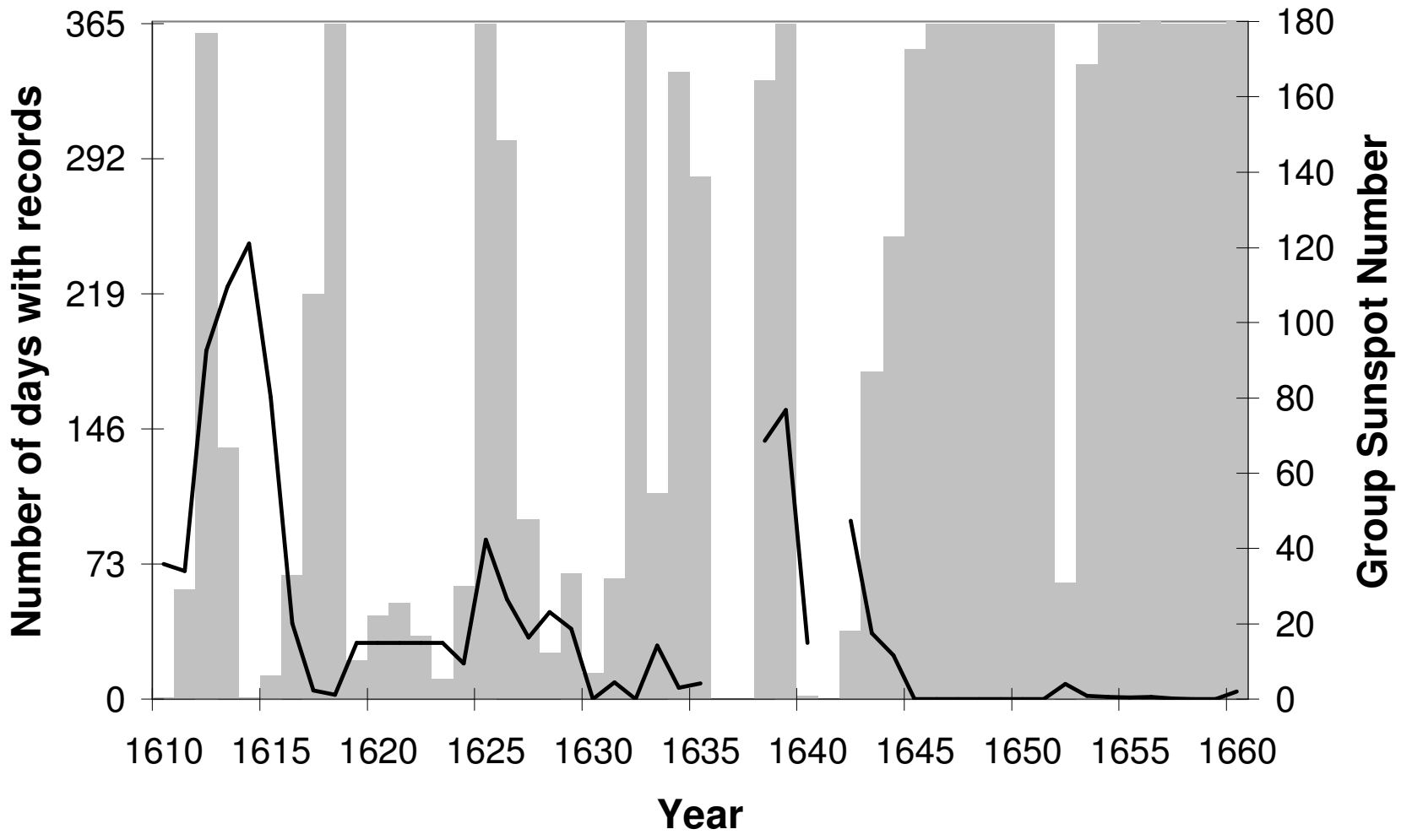

(c)

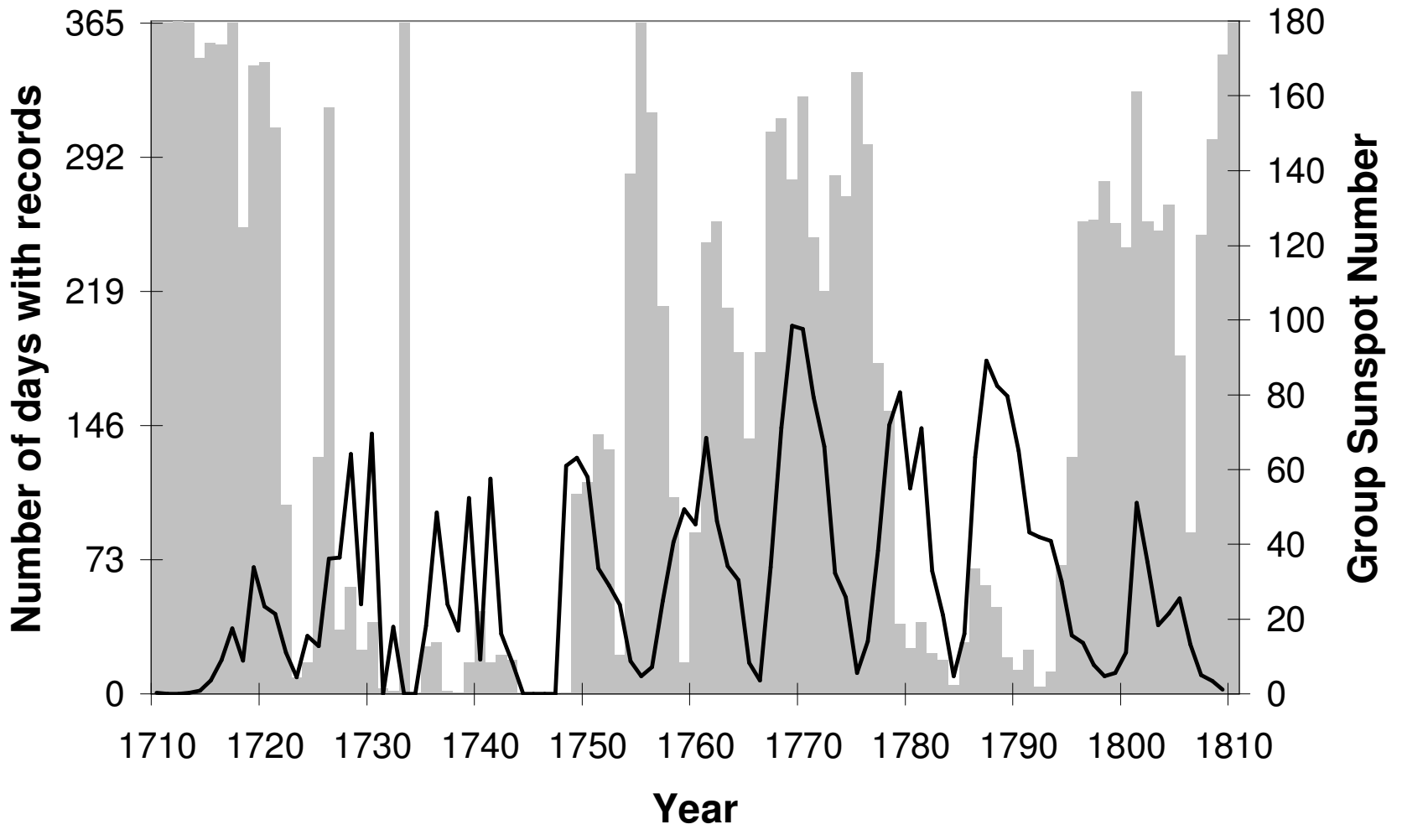

Figure 3. Number of days with records (grey bars) and $R_G$ (lines) from HS98 during (a) the entire period, (b) first half of the 17th century , and (c) 18th century.

(a)

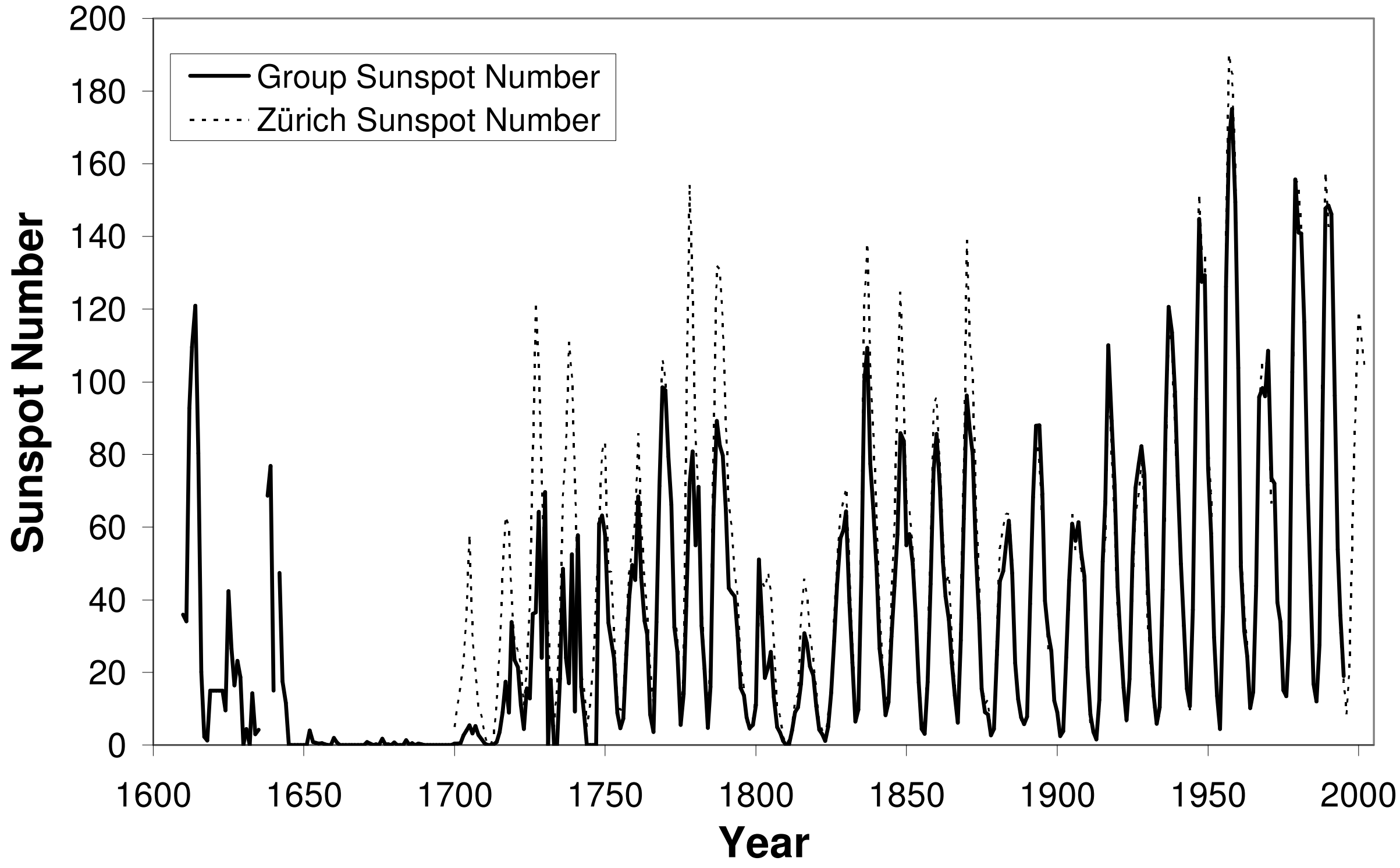

(b)

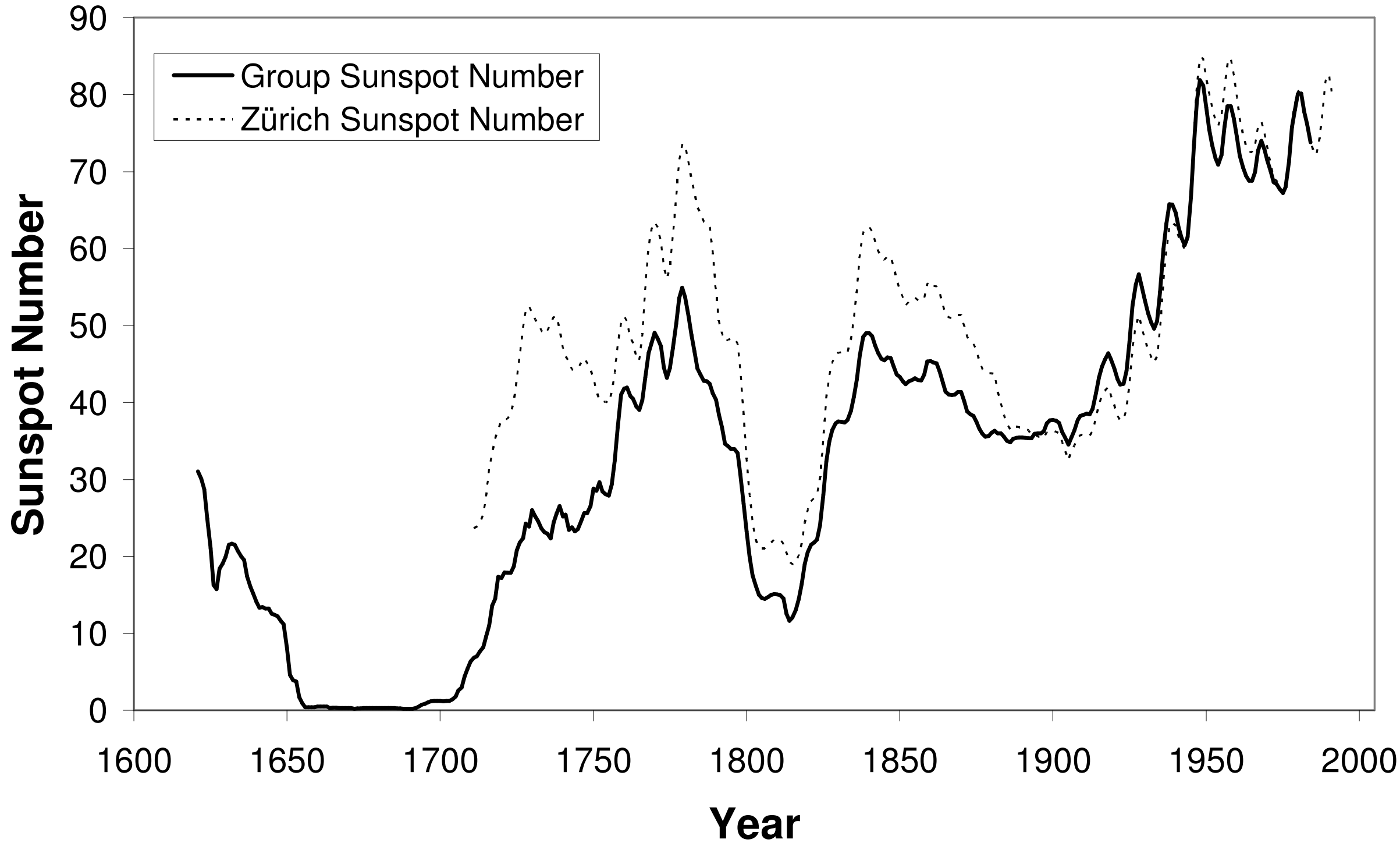

Figure 4. (a) Annual Group and Wolf Sunspot Numbers, and (b) 23-year moving averages of annual Group and Wolf Sunspot Numbers.

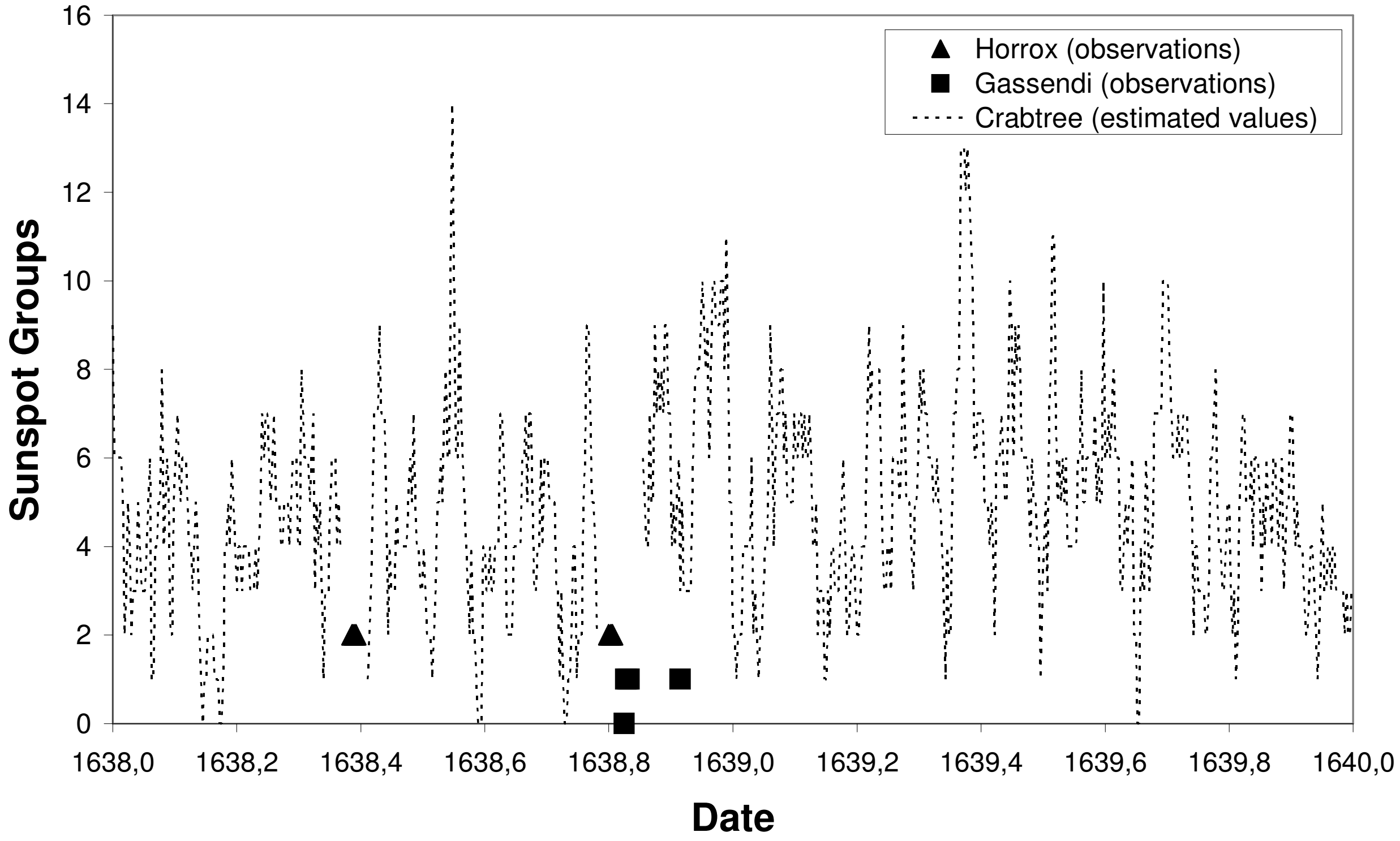

Figure 5. A comparison between the Sunspot Group values from observations (by Horrox and Gassendi) and the estimate made by HS98 from the Crabtree letter.

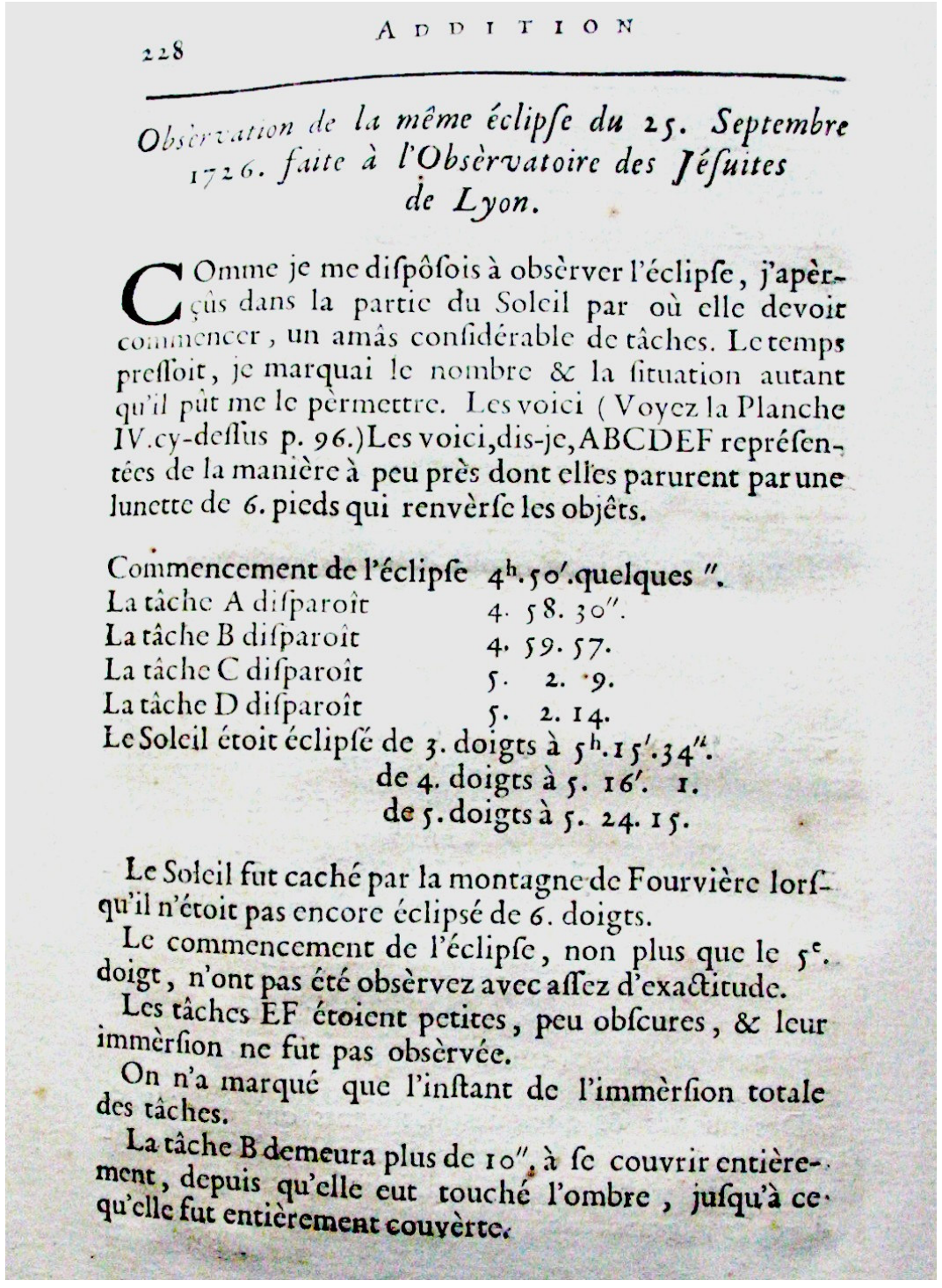

228 ADDITION

*Obsèrvation de la même éclipſe du 25. Septembre 1726. faite à l'Obsèrvatoire des Jéſuites de Lyon.*

COmme je me diſpôſois à obsèrver l'éclipſe, j'apèrçûs dans la partie du Soleil par où elle devoit commencer, un amâs conſidérable de tâches. Le temps preſſoit, je marquai le nombre & la ſituation autant qu'il pût me le pèrmettre. Les voici ( Voyez la Planche IV. cy-deſſus p. 96.) Les voici, dis-je, ABCDEF repréſentées de la manière à peu près dont elles parurent par une lunette de 6. pieds qui renvèrſe les objêts.

| | |
|---|---|
| Commencement de l'éclipſe | 4h. 50'. quelques ". |
| La tâche A diſparoît | 4. 58. 30". |
| La tâche B diſparoît | 4. 59. 57. |
| La tâche C diſparoît | 5. 2. 9. |
| La tâche D diſparoît | 5. 2. 14. |
| Le Soleil étoit éclipſé de 3. doigts à | 5h. 15'. 34". |
| de 4. doigts à | 5. 16'. 1. |
| de 5. doigts à | 5. 24. 15. |

Le Soleil fut caché par la montagne de Fourvière lorſqu'il n'étoit pas encore éclipsé de 6. doigts.

Le commencement de l'éclipſe, non plus que le 5e. doigt, n'ont pas été obsèrvez avec aſſez d'exactitude.

Les tâches EF étoient petites, peu obſcures, & leur immèrſion ne fût pas obsèrvée.

On n'a marqué que l'inſtant de l'immèrſion totale des tâches.

La tâche B demeura plus de 10", à ſe couvrir entièrement, depuis qu'elle eut touché l'ombre, juſqu'à ce qu'elle fut entièrement couvèrte.

(a)

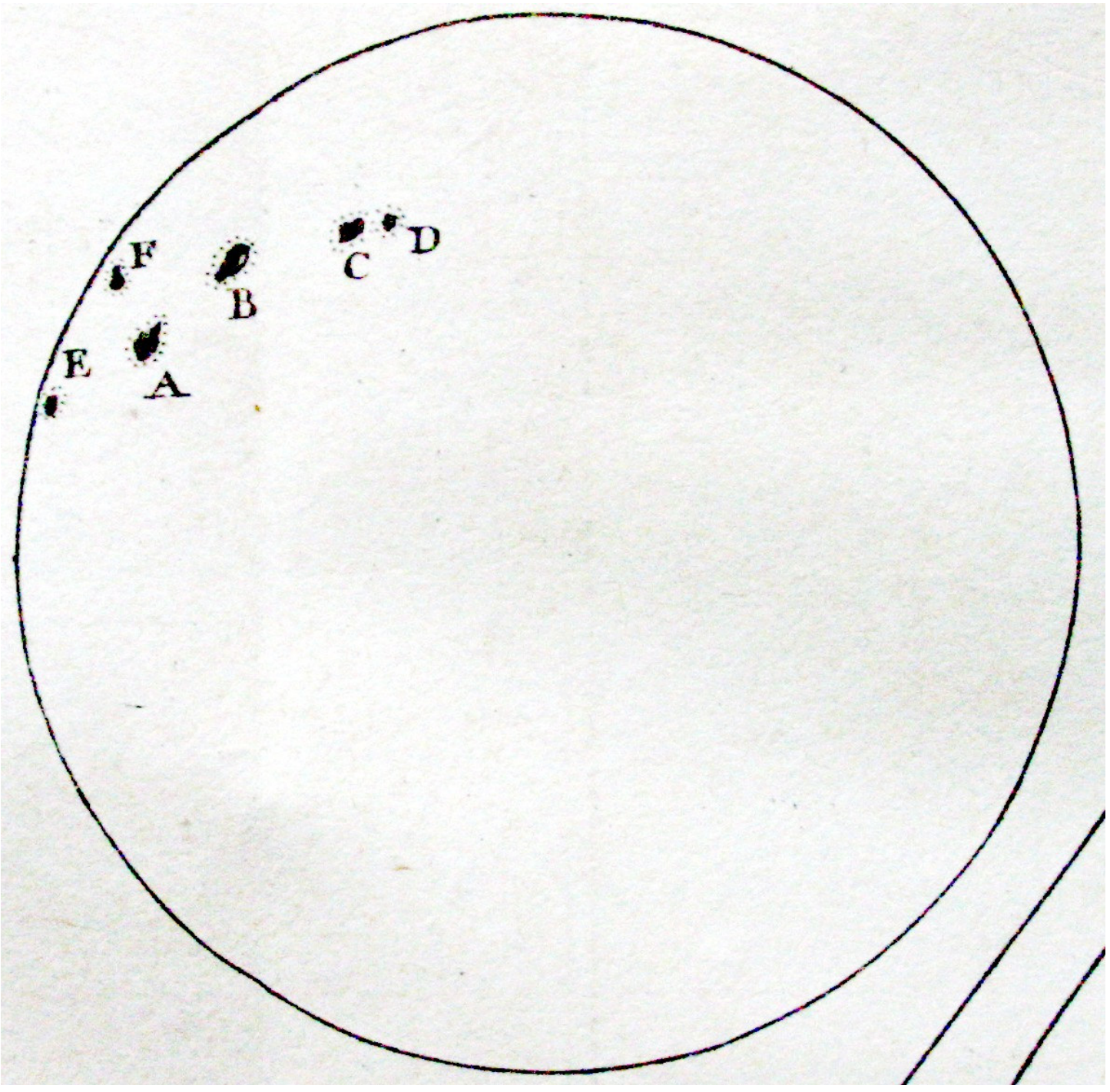

(b)

Figure 6. (a) Report by Jesuits of the solar eclipse on 25 September 1726. (b) Drawing of the sunspots on 25 September 1726. Both illustrations come from Souciet (1729).